\documentclass[prd,aps,twocolumn,nofootinbib,superscriptaddress,floatfix,preprintnumbers]{revtex4-1}

\usepackage{appendix}
\usepackage{color}
\usepackage{amsmath}
\usepackage{graphicx}
\usepackage{esint}
\usepackage{comment}
\usepackage{slashed}
\usepackage{amssymb}
\usepackage{pifont}
\usepackage[colorlinks=true,citecolor=blue, linkcolor=blue, urlcolor=blue]{hyperref}

\usepackage{slashed}

\usepackage{notoccite} 

\begin{document}

\title{Probing Neutrinophilic Scalars in Muon Decays}

\author{Yongchao Zhang}
\email{zhangyongchao@seu.edu.cn}
\affiliation{School of Physics, Southeast University, Nanjing 211189, China}
\affiliation{Center for High Energy Physics, Peking University, Beijing 100871, China}

\author{Zhong Zhang}
\email{Corresponding author, zhong.zhang.19@alumni.ucl.ac.uk}
\affiliation{School of Physics, Southeast University, Nanjing 211189, China}

\begin{abstract}

The non-standard self-interactions of neutrinos could be induced by (light) scalars that couple primarily to neutrinos. In this work, we investigate the effects of neutrinophilic scalars $\phi$ on muon decays at the tree, 1-loop and 2-loop levels, considering both Dirac and Majorana neutrinos. 
The most phenomenologically interesting processes are the 1-loop decay $\mu \to e \phi \phi$ and $\phi$-induced decay $\mu \to e \gamma$ at the 2-loop order from the same coupling $g_{\rm eff}$, 
and the 1-loop corrections of $\phi$ to the SM muon decay and the four-body muon decays $\mu^- \to e^- \nu_\mu \nu_\mu \phi,\, e^- \bar\nu_e \bar\nu_e \phi$, both induced by the coupling $h_{e\mu}$ of $\phi$ to Majorana neutrinos. 
The current experimental bounds on $\mu \to e \gamma$ provide the most stringent limits on the coupling $g_{\rm eff}$, up to roughly 0.064, for the scalar mass $m_\phi \gtrsim 8.7$ MeV. 
The precise $G_F$ measurements constrain the coupling $|h_{e\mu}|$ down to roughly $0.027$, with the scalar mass $m_\phi$ constrained up to roughly $5.9$ TeV. For $m_\phi \gtrsim {\cal O}({\rm GeV})$, the $G_F$ measurements are more stringent than other existing laboratory, astrophysical and cosmological constraints. Some other $\phi$-induced processes are also possible, which are, however, heavily suppressed by the tiny neutrino masses and/or the loop factor.
\end{abstract}

\maketitle
\flushbottom

\section{Introduction}
\label{sec:intro}

The observations of neutrino oscillations have established that at least two of the three neutrino mass eigenstates are massive, which provides the first clear laboratory evidence for new physics beyond the Standard Model (BSM). This has also motivated extensive studies of the non-standard interactions of neutrinos with other Standard Model (SM) fermions (see e.g. Refs.~\cite{Farzan:2017xzy,Proceedings:2019qno} for reviews).  
The self-interactions of neutrinos are also of great interest and can be mediated by a (light) scalar $\phi$~\cite{Berryman:2022hds}. The neutrinophilic scalars naturally arise in a variety of BSM scenarios which might involve lepton flavor violation (LFV). The couplings of $\phi$ to neutrinos could also be lepton number violating (LNV), depending on the lepton number carried by $\phi$. One of the well-motivated examples of such particles is the Majoron, which is the (pseudo-)Nambu-Goldstone boson associated with the breaking of global lepton number symmetry~\cite{Chikashige:1980ui,Gelmini:1980re,Schechter:1980gr}. Such scalars may also emerge 
from ultraviolet (UV)-complete models based on the seesaw mechanism~\cite{Heurtier:2016iac,deGouvea:2019qaz,Dev:2021axj}.

The rich phenomenology of neutrinophilic scalars has been investigated intensively in a wide range of laboratory, astrophysical and cosmological settings. Such scalars can have effects on muon decays~\cite{Goldman:1982bd,Santamaria:1986kg,Jahedi:2025hnu,Berryman:2018ogk,Foroughi-Abari:2026ofc,Lessa:2007up}, 
tau lepton decays~\cite{deGouvea:2019qaz,Brdar:2020nbj,Dev:2024ygx,Lessa:2007up}, charged meson decays~\cite{Barger:1981vd,Gelmini:1982rr,Glashow:1985cm,Lessa:2007up,Pasquini:2015fjv,Berryman:2018ogk,Bickendorf:2022buy,Dev:2024ygx,Dev:2025tdv}, $Z$ and $W$ boson decays~\cite{Berryman:2018ogk,deGouvea:2019qaz, Brdar:2020nbj,Dev:2024ygx,Agashe:2023itp, Agashe:2024owh,Zhang:2024meg,Foroughi-Abari:2025upe,Foroughi-Abari:2025mhj,Foroughi-Abari:2026ofc}, neutrino scattering experiments~\cite{Berryman:2018ogk,Kelly:2019wow},  (neutrinoless) double-beta decays~\cite{Vergados:1981gk,Deppisch:2020ztt,Deppisch:2020sqh,Georgi:1981pg,Doi:1985dx,Doi:1987rx,Mohapatra:1988fk,Burgess:1992dt,Burgess:1993xh,Bamert:1994hb,Hirsch:1995in,Mohapatra:2000px,Cepedello:2018zvr,Blum:2018ljv,Brune:2018sab,Boudjema:2025okq,deVries:2025hqa}. The 1-loop corrections of $\phi$ to the couplings of $Z$ with neutrinos will not only affect the measurements of the weak mixing angle in neutrino scattering experiments~\cite{Zhang:2024meg,Foroughi-Abari:2025mhj}, but also have effects on the non-standard interactions of neutrinos with matter and the neutrino flux measurements~\cite{Foroughi-Abari:2025mhj}; see Ref.~\cite{Foroughi-Abari:2026ofc} for the recent analysis of constraints from the electroweak precision data (EWPD). The (light) scalar $\phi$ can also be searched for at the Large Hadron Collider~\cite{deGouvea:2019qaz,Dev:2021axj}, future high-energy lepton  colliders~\cite{deLima:2024ohf,Adhikary:2024tvl,Liu:2024ywd} and the Forward Physics Facility~\cite{Kelly:2021mcd}.  
There are also potential constraints from cooling of the Sun, red giants and white dwarfs~\cite{Georgi:1981pg,Gelmini:1982rr}, supernova observations~\cite{Gelmini:1982rr,Goldman:1982bd,Kolb:1987qy,Manohar:1987ec,Choi:1987sd,Dicus:1988jh,Choi:1989hi,Kachelriess:2000qc,Tomas:2001dh,Farzan:2002wx,Davoudiasl:2005fd,Blennow:2008er,Zhou:2011rc,Galais:2011jh,Sher:2011mx,Heurtier:2016otg,Das:2017iuj,Dighe:2017sur,Brune:2018sab,Yang:2018yvk,Shalgar:2019rqe,Reddy:2021rln,Fiorillo:2022cdq,Chang:2022aas,Akita:2022etk,Akita:2023iwq,Telalovic:2024cot,Fiorillo:2023ytr,Fiorillo:2023cas,Fiorillo:2024upk,Suliga:2024nng}, the high-energy IceCube neutrino events~\cite{Ibe:2014pja,Ng:2014pca,Ioka:2014kca,Mazumdar:2020ibx,Bustamante:2020mep,Kelly:2018tyg,Hyde:2023eph,Doring:2023vmk,Foroughi-Abari:2025upe,Wang:2025qap,He:2025bex} and the cosmological observations~\cite{Bell:2005dr,Cyr-Racine:2013jua,Archidiacono:2013dua, Lancaster:2017ksf, Oldengott:2017fhy, Kreisch:2019yzn,Park:2019ibn, Escudero:2019gvw, RoyChoudhury:2020dmd, Brinckmann:2020bcn, Taule:2022jrz, Camarena:2024daj,Forastieri:2019cuf,Kreisch:2022zxp,He:2023oke,Camarena:2023cku,Huang:2017egl,Blinov:2019gcj,Venzor:2020ova,Grohs:2020xxd,Kamada:2015era,Li:2023puz,Li:2023kuz,Foroughi-Abari:2025mhj,Noriega:2025ulc}. Such neutrino self-interactions induced by $\phi$ have recently been studied in the context of the Hubble tension and the cosmological constraints on the sum of neutrino masses~\cite{Cyr-Racine:2013jua,Oldengott:2014qra,Lancaster:2017ksf,Oldengott:2017fhy,Escudero:2019gvw,Kreisch:2019yzn,Barenboim:2019tux,Blinov:2019gcj,RoyChoudhury:2020dmd,Brinckmann:2020bcn,Lyu:2020lps,Mazumdar:2020ibx,Das:2020xke,Huang:2021dba,RoyChoudhury:2022rva,Das:2023npl,Venzor:2023aka,Camarena:2023cku,He:2023oke,Poudou:2025qcx,He:2025jwp, Das:2025asx} (see e.g. Refs.~\cite{Berryman:2022hds,Dev:2025tdv} for more comprehensive lists of relevant constraints). If the scalar mass is above the MeV scale, the constraints are mainly from the charged meson decays and EWPD, while the mass range of $m_\phi \lesssim {\cal O}({\rm MeV})$ has been excluded by the big bang nucleosynthesis (BBN) data, unless the couplings are extremely small~\cite{Huang:2017egl,Blinov:2019gcj,Venzor:2020ova,Grohs:2020xxd,Kamada:2015era,Li:2023puz,Li:2023kuz,Foroughi-Abari:2025mhj}.  

In this paper, we explore the constraints on the couplings of neutrinophilic scalars from the precise muon observations. 
The muon lifetime is one of the most precisely measured quantities in particle physics, which is used to determine the Fermi constant $G_F$. 
Together with the determinations of $G_F$ from other measurements, the precise muon data are very sensitive to various BSM scenarios, in particular the exotic muon decays that are highly suppressed or completely forbidden in the SM~\cite{Gorringe:2015cma}. The branching ratio (BR) of the rare LFV decay $\mu \to e \gamma$ is experimentally constrained up to ${\cal O}(10^{-13})$, which offers complementary constraints on the BSM contributions in some of the muon decay channels~\cite{MEGII:2018kmf}. The couplings of $\phi$ with neutrinos can induce exotic muon decays, either with $\phi$ produced in the final state of muon decay at the tree level or 1-loop order, or with $\phi$ playing the role of mediator at the 1-loop or 2-loop order. This depends largely on the neutrino flavors involved and the nature of neutrinos (whether the neutrinos are Dirac or Majorana particles). It turns out that the following processes are of particular interest.
\begin{itemize}
    \item The decay $\mu \to e \phi \phi$ at the 1-loop order (cf. the diagrams in Fig.~\ref{fig:mu2ephiphi})~\cite{Goldman:1982bd,Santamaria:1986kg,Jahedi:2025hnu} and the decay $\mu \to e \gamma$ at the 2-loop level (cf. the diagrams in Fig.~\ref{fig:muegamma}), for both Dirac and Majorana neutrinos. These two processes are induced by the same sets of couplings
    of $\phi$ with neutrinos.

    \item The 1-loop $\phi$-induced corrections to the SM muon decay $\mu \to e \nu \bar\nu$ (cf. the diagrams in Fig.~\ref{fig:3-body_decay}) and the four-body decays $\mu^- \to e^- \nu_\mu \nu_\mu \phi$ and $\mu^- \to e^- \bar\nu_e \bar\nu_e \phi$ (cf. the diagrams in Fig.~\ref{fig:4-body_decay}) induced by the same LFV coupling $h_{e\mu}$, for the case of Majorana neutrinos. There are infrared (IR) divergences in the limit of $m_\phi \to 0$ in these diagrams, which are canceled out when the corresponding widths are summed up, as a result of the Kinoshita-Lee-Nauenberg (KLN) theorem~\cite{Kinoshita:1962ur,Lee:1964is}. 
    This is very similar to the case of four-body tau decay $\tau \to \ell \nu \nu \phi$~\cite{Brdar:2020nbj} (see also the IR divergence cancellation in the decays of charged meson ${\sf M} \to \ell \nu \phi$~\cite{Bickendorf:2022buy,Dev:2024ygx,Dev:2025tdv,Pasquini:2015fjv},  $\tau \to \pi \nu \phi$~\cite{Dev:2024ygx} and $Z \to \nu \nu \phi$~\cite{Brdar:2020nbj,Dev:2024ygx,Foroughi-Abari:2025mhj,Foroughi-Abari:2026ofc,Zhang:2024meg,Foroughi-Abari:2025upe}).
\end{itemize}
There are also some other $\phi$-induced processes that are phenomenologically less interesting. For instance, the two-body muon decay $\mu \to e \phi$~\cite{Santamaria:1985xa,Erwin:2006uc,Bauer:2021mvw} and the $\phi$-induced muonium-antimuonium oscillation~\cite{Ghosh:2025oju} arise at the 1-loop and 2-loop orders, respectively (cf. Figs.~\ref{fig:mu2body} and \ref{fig:Muonium}), which are both heavily suppressed by the tiny neutrino masses. The scalar $\phi$ could also contribute to the anomalous magnetic moment of muon and induce the LFV decay $\mu \to eee$ and $\mu \to e$ conversion in nuclei at the 2-loop order (cf. Figs.~\ref{fig:muegamma}, \ref{fig:mu3e} and \ref{fig:mu-e_conv}), which are, however, highly suppressed by the loop factor~\cite{Berryman:2018ogk}.

For the $\phi$-induced decays $\mu \to e \phi \phi$ and $\mu \to e \gamma$, the experimental bound on ${\rm BR}(\mu \to e \gamma)$ provides the most stringent constraints on the couplings of $\phi$ with neutrinos for the scalar mass $m_\phi \gtrsim 8.7$ MeV, excluding the couplings up to roughly 0.064 despite the 2-loop suppression (cf. Fig.~\ref{fig:limit1}). 
Regarding the 1-loop $\phi$-induced corrections to the SM muon decay $\mu \to e \nu \bar\nu$ and the four-body decays $\mu^- \to e^- \nu_\mu \nu_\mu \phi,\, e^- \bar\nu_e \bar\nu_e \phi$, the precision $G_F$ data exclude large regions of parameter space of the scalar mass $m_{\phi}$ and the coupling $h_{e\mu}$ beyond the current charged meson decays, EWPD, astrophysical and cosmological constraints, in particular for the mass range of $m_{\phi} \gtrsim {\cal O}({\rm GeV})$. The coupling $|h_{e\mu}|$ is constrained up to roughly 0.027; if the coupling is of order one, the scalar mass $m_\phi$ can be probed up to 5.9 TeV, orders of magnitude above the muon mass (cf. Fig.~\ref{fig:limit2}). 
It is remarkable that these exotic decays induced by the coupling $h_{e\mu}$ of $\phi$ could potentially be used to alleviate the mild tension at roughly $2\sigma$ C.L. in the determinations of $G_F$ from muon decay and the global electroweak (EW) fits, if the current mild tension is confirmed by future high-precision EW data~\cite{Crivellin:2021njn}. 

The rest of this paper is organized as follows. The effective couplings of $\phi$ with neutrinos are defined in Section~\ref{sec:coupling}. The decay $\mu \to e \phi\phi$ and the $\phi$-induced $\mu\to e\gamma$ are detailed in Section~\ref{sec:decay}, and the 1-loop corrections of $\phi$ to the SM decay $\mu \to e \nu \bar\nu$ and the four-body decays $\mu \to e \nu \nu \phi,\, e \bar\nu \bar\nu \phi$ are calculated in Section~\ref{sec:tree+loop}. In Section~\ref{sec:limits}, we provide calculation details of the limits on the couplings of $\phi$ with neutrinos from the $G_F$ measurements and the experimental bound on ${\rm BR}(\mu \to e \gamma)$, and list other leading existing laboratory, astrophysical and cosmological constraints. For the sake of completeness, some of the weak limits are also collected in this section. The main results are obtained in Section~\ref{sec:results}.  We comment briefly on some less interesting $\phi$-induced muon relevant processes in Section~\ref{sec:processes:small}, and conclude in Section~\ref{sec:conclusion}.

\section{Effective couplings}
\label{sec:coupling}

In this paper, we study the couplings of $\phi$ with neutrinos in a model-independent way, agnostic of the UV-completions. For the sake of concreteness and simplicity, the scalar $\phi$ is assumed  to be a real CP-even scalar, unless otherwise specified.
The interactions of $\phi$ with neutrinos depend on the nature of neutrinos, i.e. whether the neutrinos are Dirac or Majorana fermions. In the cases of Dirac and Majorana neutrinos, the relevant effective couplings are, respectively~\cite{Gluza:1991wj,Denner:1992vza}, 
\begin{subequations}
\label{eq:Lagrangian}
\begin{align}
\label{eq:LFVLagrangian}
\mathcal{L}_{\rm D}
\ = \ & - \, g_{\alpha\beta} \, \phi \, \overline{\nu_\alpha}\nu_\beta ~+~ {\rm H.c.} \,, \\
\label{eq:LNVLagrangian}
\mathcal{L}_{\rm M} \ = \ & - \, \frac12 h_{\alpha\beta} \, \phi_{} \, \overline{\nu_\alpha^c}\nu_\beta ~+~ {\rm H.c.} \,,
\end{align}
\end{subequations}
where $\nu^c$ is the charge conjugate of the neutrino field, $\alpha,\,\beta=e,\,\mu,\,\tau$ are the neutrino flavor indices, and $g_{\alpha\beta}$ and $h_{\alpha\beta}$ are the corresponding couplings in the flavor basis. The couplings above can also be written in the neutrino mass eigenstates, which are correlated to $g$ and $h$ via, respectively, 
\begin{equation}
g' = U g U^{\dagger} \,, \quad
h' = U h U^{\sf T} \,,
\end{equation}
with $U$ being the 
Pontecorvo-Maki-Nakagawa-Sakata (PMNS) matrix for neutrino mixing. In the case of $\alpha \neq \beta$, the interactions $g_{\alpha\beta}$ and $h_{\alpha\beta}$ are of the LFV type. The couplings in Eq.~(\ref{eq:Lagrangian}) could also be LNV, depending on the lepton number carried by the scalar $\phi$ which is model dependent~\cite{Bamert:1994hb,Hirsch:1995in,Schechter:1981cv,Mohapatra:1986aw, deGouvea:2019qaz}. 
For simplicity, we will not go into such details. 
Since $\phi$ is treated as a generic scalar, its mass $m_\phi$ may not be directly correlated to the neutrino mass generation, and therefore can be considered as a free parameter~\cite{deGouvea:2019qaz,Dev:2021axj,Berryman:2022hds,Brdar:2023tmi,Heurtier:2016otg,Escudero:2019gfk}.

\section{EXOTIC MUON DECAYS}
\label{sec:decay}

In this section, we consider the 1-loop decay $\mu \to e \phi \phi$ and the $\phi$-induced 2-loop decay $\mu \to e \gamma$, which depend on the same sets of Yukawa couplings. 

\subsection{Three-body muon decay $\mu\to e\phi\phi$}
\label{sec:decay:mu2ephiphi}

\begin{figure}[t!]
\centering
\includegraphics[width=0.49\linewidth]{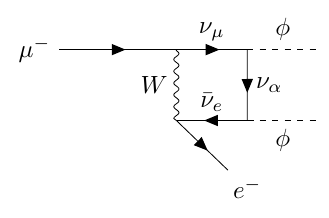}
\includegraphics[width=0.49\linewidth]{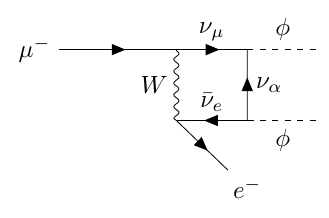}
\vspace{-20pt}
\caption{Feynman diagrams for the 1-loop three-body decay $\mu^-\to e^-\phi\phi$ for Dirac (left) or Majorana (right) neutrinos, induced by the couplings $g_{\alpha\beta}$ and $h_{\alpha\beta}$ in Eq.~(\ref{eq:Lagrangian}), respectively. }
\label{fig:mu2ephiphi}
\end{figure}

The three-body decay $\mu^-\to e^- \phi\phi$ can be induced by the (LFV) couplings $g_{e\alpha,\,\mu\alpha}$ and $h_{e\alpha,\,\mu\alpha}$ (with $\alpha = e,\, \mu,\,\tau$). The corresponding Feynman diagrams are shown in the left and right panels of Fig.~\ref{fig:mu2ephiphi}, respectively. In the limit of $m_\phi \to 0$, the experimental signals of $\mu^- \to e \phi\phi$ are quite similar to the SM muon decay $\mu \to e \nu \bar\nu$, although the electron spectra might be different. 
For Dirac neutrinos, the corresponding decay width is given by~\cite{Goldman:1982bd,Santamaria:1986kg,Jahedi:2025hnu}\footnote{In the massless neutrino limit, the matrix elements for CP-even and odd scalars differ only by an overall minus sign, therefore the corresponding decay partial widths for $\mu \to e \phi\phi$ are the same.}
\begin{equation}
    \Gamma(\mu^-\to e^-\phi\phi) \ \simeq \ \frac{G_F^2 g_{\rm eff}^4 m_\mu^5}{2^{16} \pi^7}
    \Big( 7 - 8  \log x_{\mu W}
    \Big)^2 J\left(x_{\phi\mu}\right) \,,
\label{eq:mutoephiphi}
\end{equation}
where $x_{ab} \equiv m_{a}^2/m_{b}^2$ (the muon mass $m_\mu$ and the $W$ boson mass $m_W$ are involved), the effective coupling
\begin{equation}
\label{eq:geff}
g_{\rm eff}^2 \ \equiv \ \left| \sum_\alpha g_{e\alpha}g_{\mu\alpha}^\ast \right| 
\end{equation}
incorporates all the potential contributions from the neutrino flavors $\alpha = e,\, \mu,\ \tau$ (analogously, we can also define the corresponding coupling $h_{\rm eff}$ for Majorana neutrinos), 
and the function 
\begin{equation}
    J(r) \ \equiv \ \int_{4r}^1 {\rm d} x \, (1-x)^2\sqrt{1-\frac{4r}{x}} \,.
    \label{eq:Jfunction}
\end{equation}
In the limit of $r \to 0$, $J(0)=1/3$. 
For the Majorana neutrinos, one only needs to replace $g_{\alpha\beta}$ by the corresponding coupling $h_{\alpha\beta}$ in Eq.~(\ref{eq:mutoephiphi}). 

It should be noted that the rare decay $\mu \to e \phi \phi$ can also be induced by the effective LFV couplings of $\phi$ with the charged leptons, e.g. from the effective dimension-5 operator ${\cal O}_5=(\phi^\dagger \phi) (\bar\mu e)$ and the dimension-6 operator ${\cal O}_6 = (\phi^\dagger i \overleftrightarrow{\partial_\mu} \phi) (\bar\mu \gamma^\mu e)$~\cite{Jahedi:2025hnu,Roig:2026wvo}. In this case, the scalar $\phi$ can be either real or complex, and could play the role of dark matter (DM). Such effective couplings have been proposed to alleviate the $G_F$ tensions measured in different SM processes~\cite{Roig:2026wvo}.

\subsection{Two-body muon decay $\mu\to e\gamma$}
\label{sec:decay:muegamma}

\begin{figure}[t!]
\centering
\includegraphics[width=0.49\linewidth]{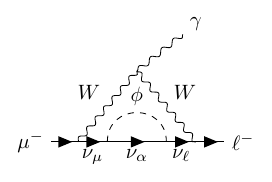}
\includegraphics[width=0.49\linewidth]{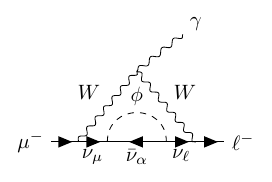}
\vspace{-20pt}
\caption{Feynman diagrams for the 2-loop two-body decay $\mu^-\to e^-\gamma$ ($\ell = e$ in the diagrams) and the anomalous magnetic moment of muon ($\ell = \mu$ in the diagrams) for Dirac (left) or Majorana (right) neutrinos, induced by the couplings $g_{\alpha\beta}$ and $h_{\alpha\beta}$ of $\phi$ in Eq.~(\ref{eq:Lagrangian}), respectively.}
\label{fig:muegamma}
\end{figure}

The couplings of $\phi$ to neutrinos contribute to the radiative LFV decay $\mu\to e\gamma$ at the 2-loop order. The corresponding Feynman diagrams for Dirac and Majorana neutrinos are shown in the left and right panels of Fig.~\ref{fig:muegamma}, respectively. In these diagrams, the scalar $\phi$ induces 1-loop corrections to the neutrino propagator in the $W$--neutrino loop, with the photon emitted from the $W$ boson. For the case of Dirac neutrinos, neglecting the tiny neutrino masses, the contribution to the BR can be parametrically estimated as
\begin{equation}
{\rm BR}(\mu\to e\gamma)
    \ = \
    \frac{3\alpha_{\rm em}}{32\pi}
    \frac{g_{\rm eff}^{4}}{(16\pi^2)^2}
    \left|F_\gamma(x_{\phi W})\right|^2 \,,
\label{eq:muegamma}
\end{equation}
where $\alpha_{\rm em}$ is the fine-structure constant, and $F_\gamma$ is the dimensionless 2-loop function. It should be noted that both the decays $\mu \to e \phi \phi$ and $\mu \to e \gamma$ are induced by the same coupling of $g_{\rm eff}$ in Eq.~(\ref{eq:geff}).

The precise form of the loop function requires a complete evaluation of the 2-loop diagrams, which is beyond the main scope of this work. In the limit of $m_\phi\ll m_W$, it is a good approximation that $F_\gamma (0) \sim 1$.
Therefore, Eq.~(\ref{eq:muegamma}) should be understood only as a parametric estimate. As in the case of $\mu \to e \phi\phi$, for the decay $\mu \to e \gamma$ involving Majorana neutrinos, one only needs to replace $g_{\alpha\beta}$ by the corresponding coupling $h_{\alpha\beta}$ in Eq.~(\ref{eq:muegamma}). 
The existing constraints on $g_{\rm eff}$ are detailed in Section~\ref{sec:limits}, with the results presented in  Section~\ref{subsec:limit:geff}, which apply equally to both the Dirac and Majorana neutrino cases here unless otherwise specified.

\section{Effects on $\mu\to e\nu\bar{\nu}$}
\label{sec:tree+loop}

In this section, we investigate the 1-loop contribution of $\phi$ to the SM muon decay $\mu\to e\nu\bar{\nu}$ and the tree-level four-body decays $\mu^- \to e^- \nu_\mu \nu_\mu \phi,\, e^- \bar\nu_e \bar\nu_e \phi$. As detailed below, there are IR divergences in these two processes in the limit of $m_\phi \to 0$, which cancel out with each other and lead to physical results.

\subsection{1-loop corrections to $\mu\to e\nu\bar{\nu}$}
\label{sec:4.1}

\begin{figure}[!t] 
\centering
\includegraphics[width=0.49\linewidth]{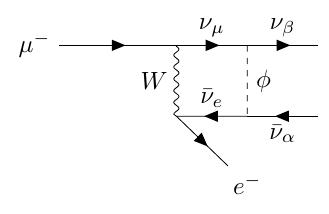}
\includegraphics[width=0.49\linewidth]{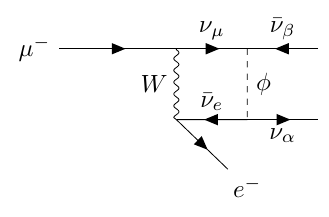}
\vspace{-20pt}
\caption{Feynman diagrams for the 1-loop decay $\mu^- \to e^- \nu_\beta \bar\nu_\alpha$ for Dirac neutrinos (left) and the decay $\mu^- \to e^- \nu_\alpha \bar\nu_\beta$ for Majorana neutrinos (right), induced by the couplings $g_{e\alpha,\,\mu\beta}$ and $h_{e\alpha,\,\mu\beta}$ (with $\alpha,\, \beta = e,\, \mu,\, \tau$) of $\phi$ in Eq.~(\ref{eq:Lagrangian}), respectively. See text for more details.}
\label{fig:3-body_decay}
\end{figure}

One of the most sensitive probes of the couplings of $\phi$ to neutrinos is provided by the  virtual corrections of $\phi$ to the ordinary muon decay $\mu^-\to e^- \nu_\mu \bar{\nu}_e$. The corresponding Feynman diagrams depend on whether the neutrinos are Dirac or Majorana particles as well as the flavor structures of the couplings $g_{\alpha\beta}$ and $h_{\alpha\beta}$ involved. 
The Feynman diagrams for the Dirac and Majorana neutrinos are shown in the left and right panels of Fig.~\ref{fig:3-body_decay}, respectively. In both cases, the relevant couplings are $g_{e\alpha,\,\mu\beta}$ and $h_{e\alpha,\,\mu\beta}$ (with $\alpha,\, \beta = e,\, \mu,\, \tau$), respectively, which can be lepton flavor conserving (LFC) or LFV. In some special cases, the neutrinos in the final state of the 1-loop diagrams are the same as the SM process $\mu^- \to e^- \bar\nu_e \nu_\mu$:\footnote{As mentioned in Ref.~\cite{Foroughi-Abari:2026ofc}, for the case of Majorana neutrinos,  the 1-loop Feynman diagram induced by the couplings $h_{ee}$ and $h_{\mu\mu}$ does not interfere with the SM process, as the (anti)neutrino flavors in the final state are different. However, it seems that the $h_{e\mu}$ coupling case was not considered in  Ref.~\cite{Foroughi-Abari:2026ofc}.}
\begin{itemize}
    \item The Dirac neutrinos, with the LFC couplings $g_{ee}$ and $g_{\mu\mu}$ (with $\alpha = e$ and $\beta = \mu$ in the left panel of Fig.~\ref{fig:3-body_decay}).
    \item The Majorana neutrinos, with the LFV coupling $h_{e\mu}$ (with $\alpha = \mu$ and $\beta = e$ in the right panel of Fig.~\ref{fig:3-body_decay}).
\end{itemize}
In these two cases, the 1-loop diagram induced by the scalar $\phi$ interferes with the tree-level SM process. The second case above is particularly of great interest, as the IR divergence from the interference of this diagram with the tree-level SM amplitude cancels out with the four-body decays $\mu^- \to e^- \nu_\mu \nu_\mu \phi,\, e^- \bar\nu_e \bar\nu_e \phi$ in Section~\ref{subsec:4-body_decay} below, which leads to physical results in the limit of $m_\phi \to 0$.\footnote{It should be noted that the IR behaviors are different for final state particles with CP-even and odd scalars~\cite{Barger:1981vd,Lessa:2007up,Bauer:2021mvw,Dev:2024ygx}.} Therefore, on the 1-loop corrections to the SM muon decay $\mu^- \to e^- \nu_\mu \bar\nu_e$, we will focus primarily on the $h_{e\mu}$ scenario throughout this work. Similar effects of the 1-loop corrections to the tau lepton decay $\tau \to \ell \nu \bar\nu$ (with $\ell = e,\,\mu$) have been studied in great detail in Ref.~\cite{Brdar:2020nbj}.

For the $\phi$-induced 1-loop decay $\mu^-(p) \to  e^-(p_1)+ \bar{\nu}_\beta(p_2) + \nu_\alpha(p_3)$, the loop integral can be written as 
\begin{widetext}
\begin{eqnarray}
\mathcal{I}_{e\nu\nu} &=&
\frac{g_{\mu\nu}}{2} \int {\rm d}^4q \frac{i(\slashed{q} + m_1)
\gamma^\nu P_L
i(\slashed{q} - \slashed{p}_2 - \slashed{p}_3 + m_2)
\gamma^\mu P_L}{(q^2-m_1^2)((q-p_3)^2-m_\phi^2)((q-p_2-p_3)^2-m_2^2)((q-p)^2-m_{W}^2)} \nonumber \\
&=& C_0\left(m_{\nu_2}^2, s_{13}, m_e^2; m_2,m_\phi, m_W\right) + C_0\left(m_\mu^2, m_{\nu_3}^2, s_{13} ; m_W, m_1, m_\phi\right) \nonumber \\
&& + \left(m_1^2+m_2^2-s_{23}\right) D_0\left(m_\mu^2, m_{\nu_1}^2, m_{\nu_2}^2,m_e^2, s_{13},s_{23}, m_W,m_1,m_\phi,m_2\right) \,,
\label{eq:loop_integral}
\end{eqnarray}
\end{widetext}
where $P_L$ is the standard left-handed chirality projection operator, $m_{1,\,2}$ are the masses of the internal neutrino propagators, $m_{\nu_i}$ (with $i=1,\,2,\,3$) are the masses for the outgoing (anti)neutrinos, $q$ is the momentum running in the loop, and $s_{13} = (p-p_3)^2$ and $s_{23} = (p_2+p_3)^2$ are the kinematic factors. $C_0$ and $D_0$ are the standard Passarino-Veltman functions~\cite{Passarino:1978jh,Hahn:1998yk,Denner:1991kt,Denner:2005nn}, and can be evaluated using {\tt Package-X}~\cite{Patel:2016fam}. 
Neglecting the tiny neutrino masses, the loop integral can be approximated as, in the limit of $m_\phi \ll m_W$, 
\begin{equation}
\label{eqn:Ienunu}
\begin{aligned}
\mathcal{I}_{e\nu\nu} & \ \simeq 2\ C_0\left(0, m_\mu^2, m_\mu^2 ; 0, m_\phi, m_W\right) + {\cal O}(m_W^{-4})\\
& \ \simeq \ \frac{2}{m_W^2-m_\phi^2} \log x_{\phi W} \,.
\end{aligned}
\end{equation}
The interference of this diagram with the tree-level amplitude for the SM muon decay $\mu^- \to e^- \nu_\mu \bar\nu_e$ contributes to the muon decay width, which turns out to be 
\begin{equation}
\label{eqn:DeltaGamma:loop}
\Delta \Gamma_\mu^{\rm loop} = \frac{|h_{e\mu}|^2 (G_F^{\mu})^2m_W^2 m_\mu^5 }{192\sqrt{2}\pi^5 } \mathcal{I}_{e\nu\nu} \,,
\end{equation}
where $G_F^\mu$ is the Fermi constant determined from muon decay.  
There is an explicit logarithmic dependence on the scalar mass $m_\phi$ through the term $\log x_{\phi W} = \log(m_\phi^2/m_W^2)$, which diverges in the limit of $m_\phi \to 0$. 
This is characteristic of the IR divergence, 
and cancels out with the four-body muon decays in Section~\ref{subsec:4-body_decay} below,  leading to physical results in the limit of $m_\phi \to 0$.

\subsection{Four-body decays $\mu^- \to e^- \nu_\mu \nu_\mu \phi$ and $\mu^- \to e^- \bar\nu_e \bar\nu_e \phi$}
\label{subsec:4-body_decay}

\begin{figure}[t!]
\centering
\includegraphics[width=0.49\linewidth]{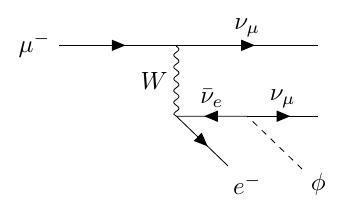}
\includegraphics[width=0.49\linewidth]{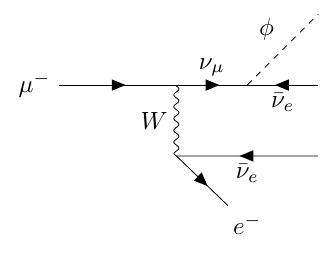}
\vspace{-20pt}
\caption{Feynman diagrams for the four-body muon decays $\mu^- \to e^- \nu_\mu \nu_\mu \phi$ (left) and $\mu^- \to e^- \bar\nu_e \bar\nu_e \phi$ (right), induced by the coupling $h_{e\mu}$ of $\phi$ with Majorana neutrinos.
}
\label{fig:4-body_decay}
\end{figure}

As just aforementioned, the logarithmic dependence of the scalar mass $m_\phi$ in Eqs.~(\ref{eqn:Ienunu}) and (\ref{eqn:DeltaGamma:loop}) in the 1-loop corrections to the SM muon decay $\mu^- \to e^- \nu_\mu \bar\nu_e$ indicates the presence of an IR divergence in the limit of $m_\phi \to 0$. A self-consistent treatment requires the inclusion of the real emission of $\phi$ in the four-body muon decays 
\begin{equation}
     \label{eqn:muon-decay:four-body}
     \mu^- \to e^- \nu_\mu \nu_\mu \phi \,, \quad
     \mu^- \to e^- \bar\nu_e \bar\nu_e \phi \,.
\end{equation}
These processes are induced by the same coupling $h_{e\mu}$ of $\phi$ as in the 1-loop diagram.  The corresponding Feynman diagrams for the decays in Eq.~(\ref{eqn:muon-decay:four-body}) are shown in Fig.~\ref{fig:4-body_decay}: in the left and right panels,  the scalar $\phi$ is emitted from the $\bar\nu_e$ and $\nu_\mu$ fermion lines, respectively. 
This is quite similar to 
the four-body tau decay $\tau \to \ell \nu \nu \phi,\, \ell \bar\nu \bar\nu \phi$~\cite{Brdar:2020nbj}, charged meson decay ${\sf M} \to \ell \nu \phi$~\cite{Bickendorf:2022buy,Dev:2024ygx,Dev:2025tdv,Pasquini:2015fjv}, three-body tau decay  $\tau \to \pi \nu \phi$~\cite{Dev:2024ygx} and the $Z$ boson decay $Z \to \nu \nu \phi,\, \bar\nu \bar\nu \phi$~\cite{Brdar:2020nbj,Dev:2024ygx,Foroughi-Abari:2025mhj,Foroughi-Abari:2026ofc,Zhang:2024meg,Foroughi-Abari:2025upe}. The cancellation of IR divergence in all these processes is a direct consequence of the KLN theorem. 
These four-body muon decays produce final states with missing energy carried by both neutrinos and the invisible scalar $\phi$, and can hardly be distinguished experimentally from the 1-loop contribution of $\phi$ to the SM process $\mu \to e \nu \bar\nu$ in the IR limit of soft $\phi$.

For the decay processes
$\mu^-(p)\rightarrow e^-(p_1) + \nu / \bar\nu (p_2)+ \nu / \bar\nu (p_3)+\phi(p_4)$ in the left and right diagrams of Fig.~\ref{fig:4-body_decay}, the reduced squared amplitudes are, respectively~\cite{deGouvea:2019qaz},
\begin{subequations}
\begin{align}
\left|\mathcal{M}_1\right|^2
= {}&
\frac{\left(p_2 \cdot p_4\right)}{\left(p_1+p_3\right)^2}
\Big[
2\left(p_2 \cdot p_4\right)
p_1 \cdot\left(p_2+p_3+p_4\right)
\nonumber\\
&
+m_\mu^2 x_{\phi \mu}\,
p_3 \cdot\left(p_1-p_2-p_4\right)
\Big] \,,
\\[1ex]
\left|\mathcal{M}_2\right|^2
= {}&
\frac{
\left(p_1 \cdot p_3\right)
+\left(p_2 \cdot p_3\right)
+\left(p_3 \cdot p_4\right)
}{
\left(p_1+p_4\right)^2
}
\nonumber\\
&\times
\Big[
2\left(p_1 \cdot p_2\right)
\left(p_1 \cdot p_4\right)
-m_\mu^2 x_{\phi\mu}
\left(p_2 \cdot p_4\right)
\Big] \,.
\end{align}
\end{subequations}
Then the corresponding partial decay width is 
\begin{eqnarray}
\Delta \Gamma_\mu^{\rm tree} 
& = & \Gamma\left(\mu^- \rightarrow e^- \nu_\mu \nu_\mu \phi\right) + \Gamma\left(\mu^- \rightarrow e^- \bar\nu_e \bar\nu_e \phi\right) \nonumber \\
& \ \simeq \ & \frac{32 G_F^2\left|h_{e\mu}\right|^2}{m_\mu} \int \mathrm{d} \Phi_4\left(\left|\mathcal{M}_1\right|^2+\left|\mathcal{M}_2\right|^2\right) \,,
\label{eq:muon_4body_decay}
\end{eqnarray}
where ${\rm d}\Phi_4$ denotes the four-body phase space in the final state. More calculation details can be found in Appendix B.2 of Ref.~\cite{deGouvea:2019qaz} (see also Ref.~\cite{Asatrian:2012tp} for details of four-body phase space integration). 
Since the two diagrams
in Fig.~\ref{fig:4-body_decay} correspond to distinct final states, their contributions are added incoherently. 
The phase space integration generates a logarithmic contribution proportional to
$\log x_{\phi\mu} $, which cancels out with the IR divergence in the virtual correction $\Delta \Gamma_\mu^{\rm loop}$ of $\phi$ to the three-body SM muon decay in Eq.~(\ref{eqn:DeltaGamma:loop}), as expected. Consequently, the inclusive decay obtained by combining the two sets of channels of $\mu\to e\nu\bar\nu$ and $\mu\to e\nu\nu\phi,\, e\bar\nu\bar\nu\phi$ is finite in the limit of $m_\phi \to 0$.

For the case of Dirac neutrinos with  the couplings $g_{ee,\,\mu\mu}$ in Section~\ref{sec:4.1}, the corresponding four-body muon decay $\mu^- \to e^- \nu_\mu \bar\nu_e \phi$ involve either $g_{ee}$ or $g_{\mu\mu}$. Therefore, the IR divergences in these diagrams  can in general not be canceled out, unless in the special case of $g_{ee} = g_{\mu\mu}$. In the following sections, we will not consider this case any more, but focus mainly on the $h_{e\mu}$ case above.

\section{Experimental constraints}
\label{sec:limits}

In this section, we collect all the existing limits on the coupling $g_{\rm eff}$ in the decays $\mu \to e \phi\phi$ and $\mu \to e \gamma$ and the coupling $|h_{e\mu}|$ for the 1-loop corrections of $\phi$ to SM muon decay and the four-body decays $\mu^- \to e^- \nu_\mu \nu_\mu \phi,\, e^- \bar\nu_e \bar\nu_e \phi$. The predominant constraints on $|h_{e\mu}|$ are from the precise measurements of the Fermi constant $G_F$  (Section~\ref{sec:GF}), and the most stringent limits on $g_{\rm eff}$ are from the current experimental bound on ${\rm BR} (\mu \to e \gamma)$ (Section~\ref{sec:muegamma:constraint}). 
Following closely Refs.~\cite{Dev:2025tdv,Foroughi-Abari:2026ofc}, there are also limits from the rare charged meson decay ${\sf M}^\pm \to \ell^\pm \nu \phi$ (Section~\ref{sec:meson}), EWPD~(Section~\ref{sec:EWPD}), the high-energy IceCube neutrino events (Section~\ref{sec:IceCube}) and the cosmic microwave background (CMB) and BBN data (Section~\ref{sec:cosmic}). The weak and irrelevant constraints are collected in Section~\ref{sec:weaklimits}. It should be noted that the constraints in this section on $g_{\rm eff}$ for Dirac neutrinos apply also to the case of Majorana neutrinos, unless otherwise specified.

\subsection{$G_F$ measurements}
\label{sec:GF}

The precisely measured muon lifetime is conventionally used to determine \(G_F^\mu\) under the assumption that the muon decay is a pure SM process~\cite{Erler:2004cx}, and the corresponding value is~\cite{Eberhart:2026klz} 
\begin{equation}
\label{eqn:value:GFmu}
G_{F}^{\mu} \ = \  1.166 378 59 (59) \times 10^{-5}\; {\rm GeV}^{-2} \,.
\end{equation}
In the presence of the \(\phi\)-induced corrections to the SM muon decay and the exotic muon decays we are considering, \(G_F^\mu\) should instead be interpreted as the experimentally inferred Fermi constant, and the comparison with other independent determinations of $G_F$ provides constraints on the couplings of $\phi$ to neutrinos. 
The global EW fit and CKM unitarity are the most important measurements for our purpose here, and the corresponding values are, respectively~\cite{Crivellin:2021njn,ParticleDataGroup:2026mpi}, 
\begin{subequations}
\begin{align}
\label{eqn:value:GFEW}
G_{F}^{\rm EW} \ = \ & 1.16716(39) \times 10^{-5} \; {\rm GeV}^{-2} \,, \\
G_{F}^{\rm CKM} \ = \ & 1.16541(43) \times 10^{-5} \; {\rm GeV}^{-2} \,, 
\end{align}
\end{subequations}
which depend substantially on the datasets and theoretical inputs used~\cite{Mildner:2024wbl,Merino:2026qxv}. 
It is remarkable that the value of $G_F$ from the global EW fit is larger than $G_F^\mu$ at the C.L. of $2\sigma$, while the value from CKM measurements is smaller than $G_F^\mu$ by  $2.3\sigma$. Despite the mild tensions among the different determinations of $G_F$, these precise measurements can still be used to set limits on exotic muon decays.  
Similar phenomenological studies have been performed for the exotic decays $\mu \to e + X + X$ with $X$ being a scalar, fermion or vector particle and potentially a DM candidate~\cite{Jahedi:2025hnu,Roig:2026wvo} (see Refs.~\cite{He:2022ljo,Liang:2023yta} for the relevant high-dimensional effective operators). 
Alternatively, these tensions could be potentially induced by exotic muon decays, e.g. the processes $\mu \to e \phi \phi$, $\mu \to e \nu \nu \phi,\, e \bar\nu \bar\nu \phi$ and the 1-loop corrections of $\phi$ to the SM decay in this paper. Then one can easily figure out the preferred parameter space in the exotic decays in this paper for these mild tensions. 

Here are more details on the constraints of $G_F$ measurements on the exotic muon decays we are considering in this paper.
\begin{itemize}
    \item As the partial width $\Gamma (\mu \to e \phi \phi)$ in Eq.~(\ref{eq:mutoephiphi}) is always positive, it can be used to alleviate the tension between $G_F^\mu$ and $G_F^\text{CKM}$. Then the corresponding constraints on $m_\phi$ and $g_{\rm eff}$ can be obtained via the condition of  
\begin{equation}
   \label{eqn:condition1}
   \Gamma (\mu \to e \phi \phi) < \frac{G_F^\mu m_\mu^5}{96\pi^3} \left( G_F^\mu-G_F^\text{CKM}\right) \,.
\end{equation}
One can also use the difference of $G_F^\mu$ and $G_F^{\rm EW}$ to set limits on $g_{\rm eff}$: as $G_F^\mu$ is smaller than $G_F^{\rm EW}$,  the corresponding constraints are expected to be stronger. To be conservative, we adopt the limit in Eq.~(\ref{eqn:condition1}) above.

\item When the virtual correction $\Delta \Gamma_\mu^{\rm loop}$ to the inclusive muon decay width in Eq.~(\ref{eqn:DeltaGamma:loop}) and the tree-level four-body partial width $\Delta \Gamma_\mu^{\rm tree}$ in Eq.~(\ref{eq:muon_4body_decay}) are summed up, the contribution of $\phi$ to the muon width is negative, which will make the value of $G_F^\mu$ larger with respect to the SM case. Therefore, we take the difference of $G_F^\mu$ and $G_F^{\rm EW}$ to set limits on $m_\phi$ and $|h_{e\mu}|$, given by 
\begin{equation}
  \label{eqn:condition2}
  |\Delta \Gamma_\mu^{\rm loop} + \Delta \Gamma_\mu^{\rm tree}| < \frac{G_F^\mu m_\mu^5}{96\pi^3} \left( G_F^{\rm EW}- G_F^\mu \right) \,.
\end{equation}
As in the case of $\mu \to e \phi\phi$ above, if the difference $G_F^\mu-G_F^\text{CKM}$ is adopted, the resulting limits become more stringent.

\end{itemize}

\subsection{$\mu\to e\gamma$}
\label{sec:muegamma:constraint}

The radiative LFV decay $\mu \to e\gamma$ provides another potential probe of the couplings $g_{\rm eff}$. 
As discussed in Section~\ref{sec:decay:muegamma}, the neutrinophilic scalar $\phi$ contributes  to the LFV decay $\mu \to e \gamma$ at the 2-loop level. The current limit from MEG II is~\cite{MEGII:2018kmf}
\begin{equation}
{\rm BR}(\mu \to e \gamma)
<
1.5\times10^{-13} \,.
\end{equation}
Despite the  2-loop suppression and the fourth power dependence of ${\rm BR}(\mu \to e \gamma)$ on the coupling $g_{\rm eff}$ in Eq.~(\ref{eq:muegamma}), the resulting bounds on the coupling $g_{\rm eff}$ turn out to be the most stringent for the parameter space of interest in this paper. The ultimate sensitivity of ${\rm BR}(\mu \to e \gamma)$ at MEG II is approximately $6\times10^{-14}$, and the constraints on $g_{\rm eff}$ will be strengthened by an extra factor of 1.3~\cite{MEGII:2018kmf}.

\subsection{Meson decays}
\label{sec:meson}

Given the couplings $g$ or $h$ to neutrinos, the light scalar $\phi$ can be produced in the leptonic decays of charged mesons ${\sf M}^\pm \to \ell_\alpha^\pm + \nu_\beta + \phi$ (with $\alpha = e,\,\mu$)~\cite{Barger:1981vd,Gelmini:1982rr,Glashow:1985cm,Lessa:2007up,Pasquini:2015fjv,Berryman:2018ogk,Bickendorf:2022buy,Dev:2024ygx,Dev:2025tdv}. These three-body decays also suffer from the IR divergence in the limit of $m_\phi \to 0$. As in the muon case in Section~\ref{sec:tree+loop}, this divergence can be canceled out once the 1-loop corrections
to the SM decay ${\sf M}^\pm \to \ell^\pm \nu$ are taken into account~\cite{Bickendorf:2022buy,Dev:2024ygx,Dev:2025tdv,Pasquini:2015fjv}.  Then the uncertainties of the corresponding partial widths $\Delta \Gamma ({\sf M}^\pm \to \ell^\pm \nu)$ can be used to set limits on the couplings $|g_{\alpha\beta}|$ or $|h_{\alpha\beta}|$ of the scalar $\phi$ to neutrinos, which can be either flavor conserving ($\alpha = \beta$) or violating ($\alpha \neq \beta$). In other words, the charged meson decay data are only sensitive to the single coupling $g_{\alpha\beta}$ or $h_{\alpha\beta}$, but not directly to the combination of $g_{\rm eff}$ in Eq.~(\ref{eq:geff}) for the decay $\mu \to e \phi\phi$. 
In practice, the experimental and lattice uncertainties of the masses, lifetimes, decay constants and relevant CKM matrix elements of the charged mesons $\pi^\pm,\, K^\pm,\,D^\pm,\, B^\pm$ are combined to obtain the meson decay constraints on $|h_{e\mu}|$ in our case~\cite{ParticleDataGroup:2026mpi,fpi,FlavourLatticeAveragingGroupFLAG:2024oxs}. 
We adopt the limits given in Fig.~6 of Ref.~\cite{Dev:2025tdv}: in the electron mode the most stringent constraints are from the decays $K^\pm,\, D^\pm \to e^\pm +\nu$, while in the muon channel the most stringent limits are $\pi^\pm,\, D^\pm \to \mu^\pm +\nu$. 
The extra particle $X$ has also been searched for in charged meson decays $\pi^\pm \to e^\pm / \mu^\pm + \nu + X$  by PIENU~\cite{PIENU:2021clt} and $K^\pm \to \mu^\pm + \nu + X$ in NA62~\cite{NA62:2021bji}, with the particle $X$ decaying invisibly. However, these spectrum data cannot exclude any new parameter space of $m_\phi$ and $|h_{e\mu}|$ beyond the meson width data above~\cite{Dev:2024ygx}.

\subsection{Electroweak precision data}
\label{sec:EWPD}

The 1-loop corrections of $\phi$ to the $Z\nu\bar\nu$ and $W\ell\nu$ vertices will affect all the EW observables, e.g. the $G_F$ measurements from muon decay, the $Z$-pole observables and the weak mixing angle. As a result, the couplings $g$ and $h$ of the scalar $\phi$ can be constrained by EWPD~\cite{Foroughi-Abari:2026ofc}. The constraints in Ref.~\cite{Foroughi-Abari:2026ofc} are obtained from a global fit to the EW observables, after incorporating the $\phi$-induced 1-loop corrections to the neutrino-relevant neutral-current and charged-current interactions. The 1-loop corrections due to $\phi$ contain the
logarithmically enhanced terms proportional to
$\ln(\Lambda^2/m_\phi^2)$ (with $\Lambda$ denoting the characteristic mass scale of the UV completion), which are, however, not sensitive to the UV details. The EWPD limits on the couplings of $\phi$ depend largely on the flavor structure. For practical purposes, we adopt the limits at the $2\sigma$ C.L. in the bottom right panel in Fig.~1 of Ref.~\cite{Foroughi-Abari:2026ofc} for the flavor-universal couplings.

\subsection{IceCube limits}
\label{sec:IceCube}

The couplings of $\phi$ with neutrinos can induce the scattering of  the high-energy astrophysical neutrinos off the cosmic neutrino background (C$\nu$B), leading to absorption lines in the IceCube neutrino spectra. Therefore, the couplings $g$ and $h$ are constrained by the IceCube data~\cite{Ibe:2014pja,Ng:2014pca,Ioka:2014kca,Mazumdar:2020ibx,Bustamante:2020mep}. The observed events from the neutrino point sources (NPSs) TXS 0506+056~\cite{IceCube:2018cha} and NGC 1068~\cite{IceCube:2022der} by IceCube provide additional constraints on the couplings $g$ and $h$~\cite{Kelly:2018tyg,Hyde:2023eph,Doring:2023vmk}. The signals of the high-energy astrophysical neutrinos at IceCube depend largely on the neutrino flavors: the shower events are from the charged-current interactions of $\nu_{e,\,\tau},\; \bar\nu_{e,\,\tau}$ and the neutral-current interactions of all (anti)neutrino flavors, while the track signals are induced by the charged-current interactions of the muon (anti)neutrinos~\cite{IceCube:2016zyt}. As a result, the IceCube limits on $g$ and $h$ depend to some extent on the neutrino flavors involved~\cite{Ng:2014pca,Mazumdar:2020ibx}. The interactions of $\phi$ with neutrinos are defined in the flavor basis, while the propagation of the high-energy astrophysical neutrinos is described in the mass basis. These two sets of neutrino states are correlated via the PMNS matrix, which is subject to the neutrino mass ordering. Therefore, the IceCube limits depend also on the neutrino mass ordering, i.e. whether the three neutrinos follow the normal or inverted mass ordering~\cite{Mazumdar:2020ibx}. To the best of our knowledge, there is not yet any dedicated analysis on the couplings $g_{\rm eff}$ or $h_{e\mu}$ in this paper; for practical purposes, we adopt the limits from the IceCube High Energy Starting Events~\cite{Bustamante:2020mep} and the NPSs~\cite{,Doring:2023vmk} on the universal flavor conserving couplings $g_{\alpha\alpha}$ (with $\alpha = e,\, \mu,\, \tau$) for both the couplings  $g_{\rm eff}$ and $h_{e\mu}$. 

\subsection{BBN and CMB limits}
\label{sec:cosmic}

In the limit of large scalar mass $m_\phi$, the couplings of $\phi$ with neutrinos generate the effective self-interactions of neutrinos through the $\phi$-mediated process of  $\nu_i \bar\nu_j \rightarrow \nu_k \bar\nu_l$ (with $i,j,k,l$ being the indices for neutrino mass eigenstates). The corresponding effective interaction strength $G_{\rm eff} \simeq g^2/m_{\phi}^2$ or $h^2/m_\phi^2$ can modify the neutrino free streaming in the early Universe, and thus get constrained by the precise measurement of the matter power spectrum~\cite{Bell:2005dr,Cyr-Racine:2013jua,Archidiacono:2013dua, Lancaster:2017ksf, Oldengott:2017fhy, Kreisch:2019yzn,Park:2019ibn, Escudero:2019gvw, RoyChoudhury:2020dmd, Brinckmann:2020bcn, Taule:2022jrz, Camarena:2024daj,Forastieri:2019cuf,Kreisch:2022zxp,He:2023oke,Camarena:2023cku}. 
One should note that the neutrino scattering process $\nu_i \bar\nu_j \rightarrow \nu_k \bar\nu_l$ involves  the PMNS matrix, and as a result the CMB constraints on the neutrino self-interactions are flavor dependent~\cite{Das:2020xke,Mazumdar:2020ibx,Brinckmann:2020bcn}. Furthermore, if only one or two neutrinos have self-interactions and the other neutrinos are free streaming, the CMB limits tend to get weaker~\cite{Das:2020xke}. As a rough approximation, the simple scenario with only one non-vanishing element of $h_{e\mu}$ corresponds to the case of two neutrinos $\nu_{e,\,\mu}$ with self-interactions and the third one $\nu_\tau$ free streaming. Then we adopt the corresponding CMB limit of $G_{\rm eff}\lesssim 6.0\times10^{-5}$ MeV$^{-2}$ in Ref.~\cite{Das:2020xke}. 
For the case of $g_{\rm eff}$, we take the limit of $G_{\rm eff}\lesssim 5.6\times10^{-5}$ MeV$^{-2}$ from Ref.~\cite{Camarena:2024daj}, with the assumption of universal couplings for all the neutrinos.

If sufficiently light, the scalar $\phi$ may also potentially contribute to the cosmological effective number of neutrino species $N_{\rm eff}$, and the corresponding limit of Planck data on the scalar mass is $m_\phi > 1.3$  MeV~\cite{Huang:2017egl,Blinov:2019gcj,Venzor:2020ova,Grohs:2020xxd}. This BBN limit is not sensitive to the flavor structure of the couplings $g$ and $h$~\cite{Blinov:2019gcj}, and we take the value of $1.3$ MeV as the limit on $m_\phi$ for the $h_{e\mu}$ case. Here for simplicity we have assumed the scalar $\phi$ is real; for the case of a complex scalar, the corresponding BBN limit gets more stringent. 
The interactions of the scalar mediator $\phi$ with neutrinos are also stringently constrained by $N_{\rm eff}$~\cite{Kamada:2015era,Li:2023puz,Li:2023kuz,Foroughi-Abari:2025mhj}. With the universal couplings of $\phi$ with neutrinos, the scalar mass is constrained to be $m_{\phi} > 6$ MeV~\cite{Li:2023kuz}, which is used as the limit on $m_\phi$ for the $g_{\rm eff}$ case. To the best of our knowledge, there is not yet any dedicated study for the flavor-specific limit on $h_{e\mu}$.

\subsection{Other weak limits}
\label{sec:weaklimits}

For the sake of completeness, we collect here the weaker limits on the couplings of $\phi$ with neutrinos.

\paragraph{Spectrum analysis of Michel electrons from muon decay} 
The emission of $\phi$ in the three-body muon decay $\mu \to e \phi\phi$, the four-body muon decays  $\mu \to e \nu \bar\nu \phi,\, e\nu\nu\phi,\, e\bar\nu\bar\nu\phi$ and the 1-loop corrections to the SM muon decay could all distort to some extent the spectrum of  electrons from muon decay, which is parameterized in terms of the Michel parameters $\rho$, $\eta$, $\xi$ and $\delta$~\cite{Renga:2019mpg}. The most precise measurements of these parameters are from the TWIST experiment, with the precision of order ${\cal O}(10^{-4})$~\cite{TWIST:2011jfx,TWIST:2011aa}.
There have been studies of such effects for the four-body decays $\mu \to e \nu \bar\nu \phi$, with the scalar $\phi$ coupling to neutrinos, charged leptons $e,\,\mu$ or in the effective dimension-5 operator of the form $\phi (\bar\mu \gamma_\alpha P_L \nu_\mu) (\bar\nu_e \gamma^\alpha P_L e)$~\cite{Lessa:2007up,Budhraja:2024diy,Bickendorf:2022buy} (see also Ref.~\cite{Greljo:2025ljr}). However, the resulting constraints on the coupling $|h_{e\mu}|$ are rather weak, only up to ${\cal O} (0.1)$ for scalar mass $m_{\phi} \gtrsim {\cal O} ({\rm MeV})$. For the decay $\mu \to e \phi\phi$, the corresponding limit is further suppressed by the coupling $g_{\rm eff}$, as $\Gamma (\mu \to e \phi\phi) \propto g_{\rm eff}^4$. 
The production of DM $\chi$ from the rare muon decay $\mu \to e \chi\chi$ can also be constrained via the Michel parameters~\cite{Barducci:2026} at future muon experiments such as Mu3e~\cite{Mu3e:2020gyw} and MEG II~\cite{MEGII:2018kmf}. However, this analysis is performed in the framework of effective field theory, and cannot be directly applied to the processes in this paper.

\paragraph{Searches of the LFV decay $\mu^- \to e^- \nu_e \bar\nu_\mu$} There have been measurements of the LFV muon decay $\mu^- \to e^- \nu_e \bar\nu_\mu$, with the precision of 1.2\%~\cite{Freedman:1993kz}, which are used to set limits on exotic muon decays~\cite{Jahedi:2025hnu,Roig:2026wvo}. However, the corresponding bounds are expected to be weaker than those from the precise $G_F$ measurements in Section~\ref{sec:GF}~\cite{Roig:2026wvo}.

\paragraph{Tau, $Z$ and $W$ boson decays} 
The three-body tau lepton decays $\tau^\pm \to \pi^\pm \nu \phi,\, \pi^\pm \bar\nu \phi$ are quite similar to the meson decays above. However, in light of the poor precision of tau lepton data, the corresponding limits are much weaker~\cite{Dev:2024ygx}. The four-body tau decays $\tau^\pm \to \ell^\pm \nu_{}  \bar\nu \phi,\,  \ell^\pm \nu_{}  \nu \phi,\, \ell^\pm \bar\nu \bar\nu \phi$ are highly suppressed by the phase space~\cite{Lessa:2007up,deGouvea:2019qaz,Brdar:2020nbj}. 
The scalar $\phi$ can also induce rare gauge boson decays $Z \to \nu \bar\nu \phi,\, \nu \nu \phi,\, \bar\nu \bar\nu \phi$ and $W^\pm \to \ell^\pm \nu \phi,\, \ell^\pm \bar\nu \phi$. However, the constraints from the $Z \to {\rm inv}$ and $W^\pm \to \ell^\pm \nu$ data are also rather weak, at most $g \lesssim {\cal O}(1)$~\cite{Berryman:2018ogk,deGouvea:2019qaz, Brdar:2020nbj,Dev:2024ygx,Agashe:2023itp, Agashe:2024owh,Zhang:2024meg,Foroughi-Abari:2025upe,Foroughi-Abari:2025mhj,Foroughi-Abari:2026ofc}. 
Other weak laboratory constraints can be found e.g. in Ref.~\cite{Dev:2025tdv}. 

\paragraph{Double-beta decays} 
If neutrinos are Majorana particles, the scalar $\phi$ can be emitted in neutrinoless double-beta ($0\nu\beta\beta$) decays from the coupling $h_{ee}$ to electron neutrinos, via the process of $(Z,\, A) \to (Z+2,\,A) + e^- + e^- + \phi$, as long as the scalar mass is below the corresponding $Q$ values of the nuclear double-beta transitions~\cite{Georgi:1981pg} (see Refs.~\cite{Burgess:1992dt,Burgess:1993xh,Bamert:1994hb,Hirsch:1995in,Cepedello:2018zvr,Blum:2018ljv,Brune:2018sab,Boudjema:2025okq,deVries:2025hqa,Bolton:2026gku} for the rich phenomenological studies).  
The searches for this channel have been performed in the $0\nu\beta\beta$ decay experiments~\cite{Bernatowicz:1992ma, Arnold:2018tmo,NEMO-3:2019gwo,KamLAND-Zen:2012uen,Kharusi:2021jez,GERDA:2022ffe,CUPID-0:2022yws,CUPID:2024qnd} and PandaX-4T~\cite{PandaX:2025tls}, with the coupling $h_{ee}$ excluded up to ${\cal O}(10^{-5})$. 
However, these limits apply only to the coupling $h_{ee}$ for electron flavor, and are not relevant to the LFV coupling $h_{e\mu}$ and the coupling $g_{\alpha\beta}$ for Dirac neutrinos. 
Even if the couplings of $\phi$ to neutrinos are LFV and neutrinos are Dirac particles, e.g. in the form of Eq.~(\ref{eq:LFVLagrangian}), 
the double-beta decays can proceed via the process of $(Z,\, A) \to (Z+2,\,A) + e^- + e^- + \bar\nu_\alpha + \bar\nu_\beta$ (with $\alpha,\,\beta = e,\, \mu,\,\tau$), but the corresponding experimental two-neutrino double-beta decay limits on the coupling $g$ are rather weak~\cite{Deppisch:2020sqh,Dev:2025tdv}.

\paragraph{Supernova limits}  
The dense neutrinos play a crucial role in the explosion of core-collapse supernovae. The non-standard self-interactions of neutrinos induced by the light scalar $\phi$ will not only have potential effects on the cooling of supernova cores, but also affect the neutrino evolution, shock revival and the supernova explosion~\cite{Gelmini:1982rr,Goldman:1982bd,Kolb:1987qy,Manohar:1987ec,Choi:1987sd,Dicus:1988jh,Choi:1989hi,Bilenky:1999dn,Kachelriess:2000qc,Tomas:2001dh,Farzan:2002wx,Davoudiasl:2005fd,Blennow:2008er,Zhou:2011rc,Galais:2011jh,Sher:2011mx,Heurtier:2016otg,Das:2017iuj,Dighe:2017sur,Brune:2018sab,Yang:2018yvk,Shalgar:2019rqe,Reddy:2021rln,Fiorillo:2022cdq,Chang:2022aas,Akita:2022etk,Akita:2023iwq,Telalovic:2024cot}. The couplings of the scalar $\phi$ with neutrinos can also be measured in the future searches of the  diffuse supernova neutrino background~\cite{Akita:2022etk} and the scattering of supernova neutrinos with C$\nu$B~\cite{Kolb:1987qy,Bilenky:1999dn,Shalgar:2019rqe}. However, in the parameter space of interest in  this paper, the supernova neutrinos stream as a fireball, and the effects of $\phi$ on the supernova core are expected to be small~\cite{Fiorillo:2023ytr,Fiorillo:2023cas,Fiorillo:2024upk}.

\section{Results}
\label{sec:results}

Based on the analysis in Section~\ref{sec:limits}, the constraints on $g_{\rm eff}$ and $|h_{e\mu}|$ from the $G_F$ measurements and the experimental bounds on ${\rm BR}(\mu \to e \gamma)$ are obtained in this section, together with other existing laboratory, astrophysical and cosmological limits.

\subsection{Limits on $g_{\rm eff}$}
\label{subsec:limit:geff}

\begin{figure}[t!]
    \centering
     \includegraphics[width=0.95\linewidth]{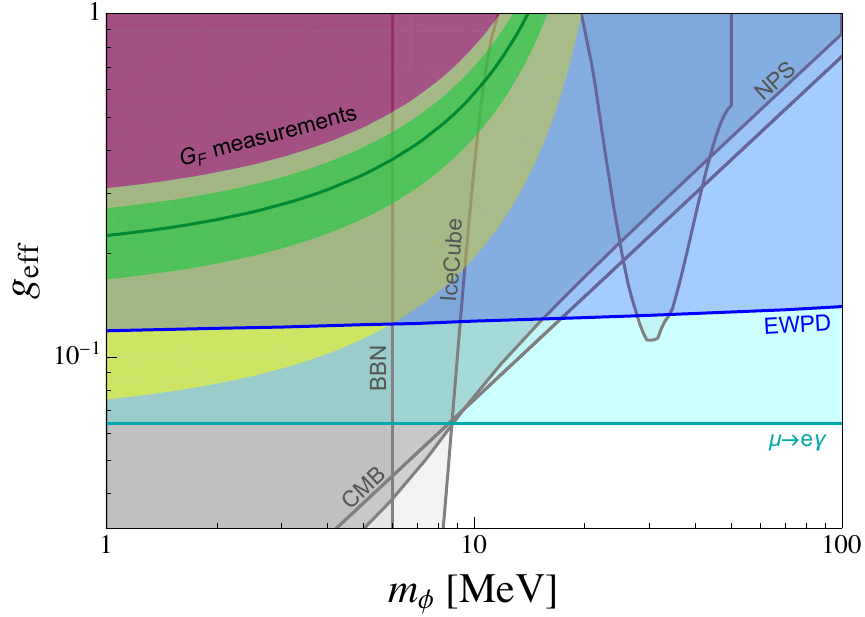}
    \caption{Constraints on the scalar mass $m_\phi$ and the effective coupling $g_{\rm eff}$ defined in Eq.~(\ref{eq:geff}). The red and cyan shaded regions are excluded by the measurements of $G_F$ and ${\rm BR}(\mu \to e \gamma)$, respectively. The dark green line denotes the central values corresponding to the mild tension between $G_F^\mu$ and $G_F^{\rm CKM}$, while the green and yellow bands indicate the corresponding preferred regions at the $1\sigma$ and $2\sigma$ C.L., respectively~\cite{Crivellin:2021njn,ParticleDataGroup:2026mpi,Eberhart:2026klz}. The other shaded regions are the existing approximate constraints from EWPD~\cite{Foroughi-Abari:2026ofc}, MEGII~\cite{MEGII:2025gzr}, IceCube~\cite{Bustamante:2020mep}, NPSs~\cite{Doring:2023vmk},  CMB~\cite{Camarena:2024daj} and BBN~\cite{Li:2023puz}. See text for more details. } 
    \label{fig:limit1}
\end{figure}

Requiring that the $\phi$-induced contribution to the muon width $\Gamma (\mu \to e \phi\phi)$ in Eq.~(\ref{eq:mutoephiphi}) is smaller than the difference of $G_F^\mu$ and $G_F^{\rm CKM}$ at the $2\sigma$ C.L. in Eq.~(\ref{eqn:condition1}), 
we can set limits on the scalar mass $m_\phi$ and the effective coupling $g_{\rm eff}^2$ in Eq.~(\ref{eq:geff}). The excluded parameter space of $m_{\phi}$ and $g_{\rm eff}$ is shown as the red shaded region in Fig.~\ref{fig:limit1}. 
As the partial width $\Gamma (\mu \to e \phi \phi) \propto g_{\rm eff}^4$, the corresponding constraints on the effective coupling $g_{\rm eff}$ are rather weak, with the coupling $g_{\rm eff}$ constrained only up to $0.4$ for scalar mass $m_\phi > 1$ MeV, as seen in this figure. 
As discussed in Section~\ref{sec:GF}, the channel $\mu \to e \phi\phi$ can be used to alleviate the tension between $G_F^\mu$ and $G_F^{\rm CKM}$, and the corresponding central values of $m_\phi$ and $g_{\rm eff}$ are presented as the dark green line in Fig.~\ref{fig:limit1}. The preferred regions at the $1\sigma$ and $2\sigma$ C.L. are indicated by the green and yellow bands, respectively.

The radiative LFV decay $\mu\to e\gamma$ provides a complementary probe of the effective coupling $g_{\rm eff}$. Although the $\phi$-induced contributions arise at the 2-loop level, the coupling $g_{\rm eff}$ is constrained by MEG II up to approximately 0.064, which is shown as
the cyan shaded region in Fig.~\ref{fig:limit1}. This is significantly stronger than that from the $G_F$ measurements above. 
As mentioned in Section~\ref{sec:decay:muegamma}, without a complete 2-loop calculation, the limit obtained here should be understood as a rough estimate. 

As detailed in Section~\ref{sec:limits}, the other existing constraints on $m_\phi$ and $g_{\rm eff}$ are mainly from the  EWPD~\cite{Foroughi-Abari:2026ofc}, 
the IceCube High Energy Starting Events~\cite{Bustamante:2020mep}, NPSs~\cite{Doring:2023vmk}, CMB~\cite{Camarena:2024daj} and BBN~\cite{Li:2023puz}, which are presented as the shaded regions in Fig.~\ref{fig:limit1}. 
It should be noted that these limits are obtained with the assumption of
universal flavor-conserving couplings, i.e. $g_{ee} = g_{\mu\mu} = g_{\tau\tau}$, and all the flavor-changing couplings set to be zero. In our case, there should be the non-diagonal couplings $g_{\alpha\beta}$ (with $\alpha\neq\beta$) to enable the LFV decay $\mu \to e \phi\phi$.  
Therefore, these contours cannot be interpreted as rigorous bounds on \(g_{\rm eff}\) for a general flavor structure; they are used only as approximate reference constraints on $g_{\rm eff}$.

Overlapping all the current limits above, it is clear in Fig.~\ref{fig:limit1} that the limit from $\mu \to e \gamma$ is the most stringent for the scalar mass range of $m_\phi \gtrsim 8.7$ MeV, while the $G_F$ constraint and the preferred regions associated with the mild tension between $G_F^\mu$ and $G_F^{\rm CKM}$ are excluded by other laboratory, astrophysical and cosmological limits in this figure. Such a conclusion also applies to the corresponding case with Majorana neutrinos, as mentioned in Section~\ref{sec:decay:mu2ephiphi}.

\subsection{Limits on $|h_{e\mu}|$}
\label{sec:limit2}

\begin{figure}[t!]
    \centering
    \includegraphics[width=0.95\linewidth]{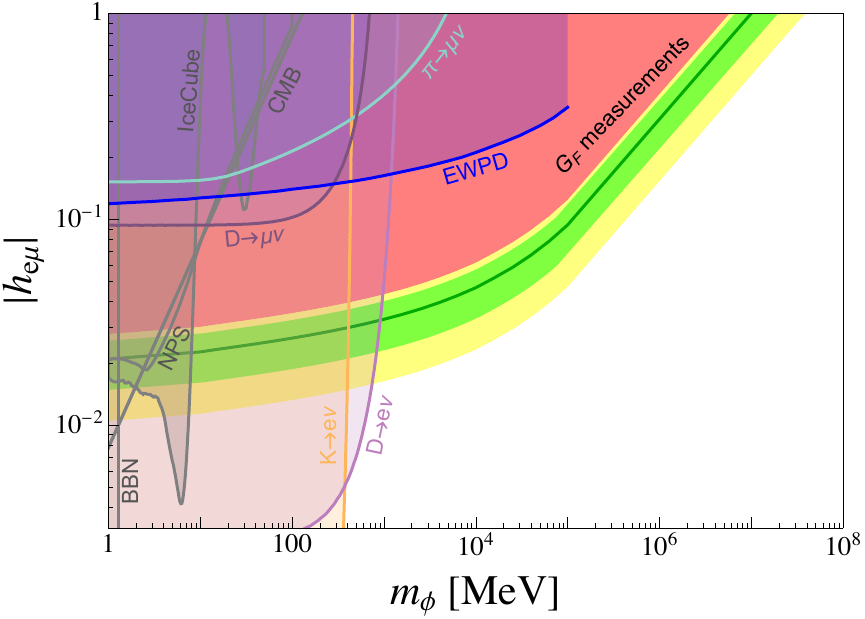}
    \caption{
    Constraints on the scalar mass $m_\phi$ and the coupling $|h_{e\mu}|$. The red shaded region is excluded by the $G_F$ measurements. The dark green line denotes the central values corresponding to the mild tension between $G_F^\mu$ and $G_F^{\rm EW}$, while the green and yellow bands indicate the corresponding preferred regions at the $1\sigma$ and $2\sigma$ C.L., respectively~\cite{Crivellin:2021njn,Eberhart:2026klz}. The other shaded regions are the existing constraints from charged meson decays ${\sf M}\to\ell\nu$~\cite{Dev:2024ygx,Dev:2025tdv} and the approximate limits from EWPD~\cite{Foroughi-Abari:2026ofc},  IceCube~\cite{Bustamante:2020mep}, NPSs~\cite{Doring:2023vmk}, CMB~\cite{Das:2020xke} and BBN~\cite{Blinov:2019gcj}. See text for more details.} 
    \label{fig:limit2}
\end{figure}

The constraint on the scalar mass $m_\phi$ and the coupling $|h_{e\mu}|$ from the precise measurements of $G_F^\mu$ and $G_F^{\rm EW}$ at $2\sigma$ C.L. in Eq.~(\ref{eqn:condition2}) is shown as the red shaded region in Fig.~\ref{fig:limit2}. The coupling $|h_{e\mu}|$ is excluded up to $0.027$  for the scalar mass $m_\phi \gtrsim 1$ MeV. In analogy to the case of charged meson decays, as a result of the 1-loop contribution of $\phi$ to the SM decay $\mu \to e \nu \bar\nu$, the scalar mass can be constrained far above the muon mass, up to 5.9 TeV with the coupling $|h_{e\mu}|\leq1$.  One should be aware that the  constraint of $G_F$ measurements on the scalar mass $m_\phi$ is obtained in the framework of effective coupling in Eq.~(\ref{eq:LNVLagrangian}). In the UV-complete models for the effective coupling $h_{e\mu}$, the corresponding limits might be to some extent different~\cite{Foroughi-Abari:2026ofc}. Nevertheless, it is expected that the constraints of $G_F$ measurements on $m_\phi$ can reach ${\cal O}({\rm TeV})$ in realistic models. 
The corresponding central values of $m_\phi$ and $|h_{e\mu}|$ for the mild tension between $G_F^\mu$  and $G_F^{\rm EW}$ are indicated by the darker green line, while the corresponding parameter space at the $1\sigma$ and $2\sigma$ C.L. are presented as the green and yellow bands, respectively.\footnote{As the value of $G_F^\mu$ and $G_F^{\rm EW}$ differ merely at the $2\sigma$ C.L. (cf Eqs.~(\ref{eqn:value:GFmu}) and (\ref{eqn:value:GFEW})), the lower boundary of the $2\sigma$ band in Fig.~\ref{fig:limit2} is taken to be actually at the $1.5\sigma$ C.L. 
}

Based on the analysis in  Section~\ref{sec:limits}, the most stringent existing laboratory limits are from the charged meson decays $K^\pm,\, D^\pm \to e^\pm +\nu$,  $\pi^\pm,\, D^\pm \to \mu^\pm +\nu$~\cite{Dev:2024ygx,Dev:2025tdv} and the EWPD~\cite{Foroughi-Abari:2026ofc}. These limits are shown as the shaded regions  in orange, light purple, cyan, dark purple and blue in Fig.~\ref{fig:limit2}, respectively. For relatively large coupling $|h_{e\mu}| \gtrsim 3\times10^{-3}$ and the scalar mass $m_\phi \lesssim$ GeV, the laboratory constraints are dominated by the decays $K,\, D \to e \nu$, while for a heavy scalar with mass $m_\phi \gtrsim$ GeV, the EWPD becomes the most stringent constraint. It should be noted that the EWPD limit adopted here is based on the assumption of universal couplings, which should be considered an approximate constraint on the LFV coupling $|h_{e\mu}|$ in our case~\cite{Foroughi-Abari:2026ofc}. 
In the parameter space of interest in the paper, the astrophysical and cosmological constraints of the IceCube High Energy Starting Events~\cite{Bustamante:2020mep}, NPSs~\cite{Doring:2023vmk}, CMB~\cite{Das:2020xke} and BBN~\cite{Blinov:2019gcj} are less constraining than the laboratory limits above, which are presented as the gray shaded regions in Fig.~\ref{fig:limit2}. 
It should be noted that these astrophysical and cosmological limits are also approximate limits on the coupling $|h_{e\mu}|$, see the discussions in Sections~\ref{sec:IceCube} and \ref{sec:cosmic}.

It is transparent in  Fig.~\ref{fig:limit2} that 
the $G_F$ measurements provide the most stringent constraints on the coupling $|h_{e\mu}|$ for the mass range of $m_\phi \gtrsim {\cal O}({\rm GeV})$, even if all other existing laboratory, astrophysical and cosmological limits are taken into account. 
Although both the $G_F$ measurements and EWPD are sensitive to the parameter space with large $m_\phi$, the $G_F$ constraint is more stringent than that from EWPD. 
In particular, at $m_\phi \simeq 1$ GeV, the limit on $|h_{e\mu}|$ is improved by a factor of 4.8 by the precision $G_F$ data with respect to that from EWPD. 

Here are some more comments on the constraints of muon decays on the neutrinophilic scalars.
\begin{itemize}
    \item The most important limits on the couplings of $\phi$ to neutrinos obtained in this paper  are mainly from the precise determinations of the Fermi constant from muon decay,  CKM unitarity and EW fit. If these data are combined with the precise measurements of the Michel parameters in muon decay and EWPD, the constraints on the couplings $g$ and $h$ are expected to be more stringent. However, this global analysis is far beyond the main scope of this paper.

    \item The precision of the couplings $g$ and $h$ of the scalar $\phi$ can be significantly improved at the next-generation high-intensity experiments. More muon events will be collected at future high-intensity experiments such as 
    Mu3e~\cite{Mu3e:2020gyw}, 
    Mu2e~\cite{Mu2e:2014fns}, COMET~\cite{COMET:2018auw}, MEG II~\cite{MEGII:2018kmf}, and MACE~\cite{Bai:2024skk}. The unprecedented precision at these experiments can not only be used to directly search for the exotic processes induced by $\phi$, but also help improve the constraining power of the Michel parameters~\cite{Barducci:2026}. Furthermore, the sensitivities of EWPD to new physics will also be significantly improved by the Large Hadron Collider, future high-energy lepton colliders and the high-intensity experiments~\cite{Crivellin:2021njn}.
\end{itemize}

\section{Some other processes induced by $\phi$}
\label{sec:processes:small}

In this section, we collect the muon relevant processes induced by $\phi$, that are less phenomenologically interesting as a result of  the suppression by the tiny neutrino masses and/or the loop factor.

\subsection{$\mu \to e \phi$}

\begin{figure}[t!]
\centering
\includegraphics[width=0.55\linewidth]{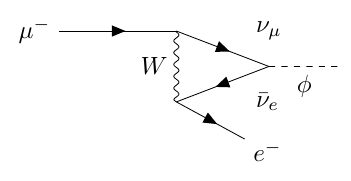}
\vspace{-10pt}
\caption{Feynman diagrams for the 1-loop $\mu^-\to e^-\phi$, induced by the coupling $g_{e\mu}$ in Eq.~(\ref{eq:LFVLagrangian}).}
\label{fig:mu2body}
\end{figure}

In the case of Dirac neutrinos, one of the simplest channels involving $\phi$ is the radiative two-body muon decay $\mu^- \to e^- \phi$, induced by the LFV coupling $g_{e\mu}$. This can be easily generalized to the case of tau decay $\tau \to \ell \phi$ (with $\ell = e,\,\mu$). The corresponding Feynman diagram is shown in Fig.~\ref{fig:mu2body},  
and the corresponding partial width reads~\cite{Santamaria:1985xa,Erwin:2006uc,Bauer:2021mvw}: 
\begin{equation}
    \Gamma\left(\mu^{-} \rightarrow e^-\phi\right) \simeq \frac{|g_{e\mu}|^2G_F^2 m_\mu^3}{2^9 \pi^5 }\left(1- x_{\phi\mu} \right)^2
    |m_{e\mu}|^2
    \label{eq:mutoephi}
\end{equation}
where the effective neutrino mass (squared) parameter is defined as 
\begin{equation}
m_{\alpha\beta}^{(2)}\equiv \sum_i m_{\nu_i}^{(2)}
U_{\alpha i}^\ast U_{\beta i} \,.
\end{equation}
It is apparent that this decay channel is highly suppressed by the tiny neutrino masses, and is therefore less phenomenologically interesting. 
The relevant experimental searches of the two-body decay $\mu \to e + X$ (with $X$ being an invisible particle) have been performed by TWIST~\cite{TWIST:2014ymv}, PIENU \cite{PIENU:2020loi} and M20 at TRIUMF \cite{Collar:2023xla} for different mass scales. 

\subsection{Muonium-antimuonium oscillation}

\begin{figure}[!t]
\centering
\includegraphics[width=0.6\linewidth]{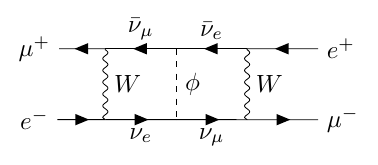} \vspace{-10pt} \\
\includegraphics[width=0.6\linewidth]{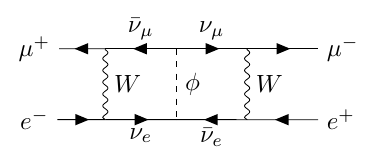}
 \vspace{-10pt}
\caption{Feynman diagrams for muonium--antimuonium oscillation induced by $\phi$ at the 2-loop order via the couplings $g_{e\mu}$ and $h_{ee,\,\mu\mu}$ in Eq.~(\ref{eq:Lagrangian}) for Dirac (top) or Majorana (bottom) neutrinos, respectively.
}
\label{fig:Muonium}
\end{figure}

Muonium is a bound state composed of an antimuon and an electron, i.e. ${\rm Mu} = (\mu^+ e^-)$. The muonium--antimuonium (${\rm Mu}$-$\overline{\rm Mu}$) oscillation is a clear signature of LFV from BSM scenarios. The process has attracted renewed interest in light of the proposed next-generation experiments such as MACE, which aims to significantly improve the oscillation sensitivity~\cite{Bai:2024skk}. For both Dirac and Majorana neutrinos, the ${\rm Mu}$-$\overline{\rm Mu}$ oscillation can be induced by the (LFV) couplings of $\phi$ with neutrinos at the 2-loop order. The corresponding Feynman diagrams are shown in Fig.~\ref{fig:Muonium}: the top diagram is induced by the LFV coupling $g_{e\mu}$ for Dirac neutrinos, while the bottom one arises from the couplings $h_{ee} h_{\mu\mu}$ of $\phi$ with Majorana neutrinos. 
The ${\rm Mu}$-$\overline{\rm Mu}$ oscillation probability  due to the top and bottom diagrams in Fig.~\ref{fig:Muonium} can be approximated by, respectively~\cite{Ghosh:2025oju}, 
\begin{subequations}
\begin{align}
P_{\rm D}({\rm Mu} \rightarrow \overline{{\rm Mu}})
\ \sim \ & \frac{ 4608 |g_{e\mu}|^4}{(4\pi)^2 (m_\mu a_0)^6}
\left| \frac{m_{e\mu}^2}{m_\mu^2} \right|^2
\log^2 x_{\phi W} \,, \\
P_{\rm M}({\rm Mu} \rightarrow \overline{{\rm Mu}})
\ \sim \ & \frac{4608 |h_{ee}h_{\mu\mu}|^2}{(4\pi)^2(m_\mu a_0)^6}
\left| \frac{m_{ee}m_{\mu\mu}}{m_\mu^2} \right|^2 \log^2 x_{\phi W} \,,
\end{align}
\end{subequations}
where $a_0\equiv(m_e\alpha_{\rm em})^{-1}$ is the Bohr radius. In both cases, the oscillation probability is suppressed not only by the loop factor, but also, more importantly, by the tiny neutrino masses via $m_{\nu_i}^2/m_\mu^2 \lesssim {\cal O}(10^{-18})$. As a result, the expected oscillation rates due to the couplings of $\phi$ are extremely small. 

\subsection{The anomalous magnetic moment of muon}

The couplings of $\phi$ to neutrinos can also contribute to the muon anomalous magnetic moment $a_\mu$ at the 2-loop order, and the dominant contributions are via the $\phi$-induced corrections to the neutrino propagator in the  $W$-$\nu$ loop. The corresponding Feynman diagrams are shown in Fig.~\ref{fig:muegamma}, with the left and right panels for the Dirac and Majorana neutrino cases, induced by the couplings $g_{\mu \alpha}$ and $h_{\mu \alpha}$ (with $\alpha = e,\, \mu,\, \tau$), respectively.  
Neglecting the tiny neutrino masses, the $\phi$-induced contribution for the case of Dirac neutrinos can be parametrically estimated as
\begin{equation}
\label{eqn:g-2}
\Delta a_\mu
\sim \frac{G_F m_\mu^2}{64\sqrt{2}\pi^4}
 \sum_\alpha |g_{\mu \alpha}|^2 \,,
\end{equation}
where we have assumed $m_\phi \ll m_W$. 
For Majorana neutrinos, one only needs to replace the coupling $g_{\mu\alpha}$ by $h_{\mu \alpha}$ in Eq.~(\ref{eqn:g-2}). 
The precise form of $\Delta a_\mu$ depends on the complete 2-loop calculations, which is beyond the main scope of this work. Nevertheless, 
the $\phi$-induced contributions are highly suppressed by the loop factor, with $\Delta a_\mu^\phi \sim 10^{-11} \sum_\alpha |g_{\mu\alpha}|^2$.  The current precision of $a_\mu$ is roughly $10^{-10}$, and the corresponding constraints on the couplings $g_{\mu\alpha}$ and $h_{\mu\alpha}$ are therefore rather weak~\cite{Muong-2:2025xyk}. The precision of the anomalous magnetic moment of electron $a_e$ is much higher than that for muon, reaching ${\cal O}(10^{-13})$, and the resulting constraints on the corresponding couplings $\sum_\alpha |g_{e\alpha}|^2$ are expected to be more stringent.

\subsection{$\mu \to eee$}

\begin{figure}[t!]
\centering
\includegraphics[width=0.55\linewidth]{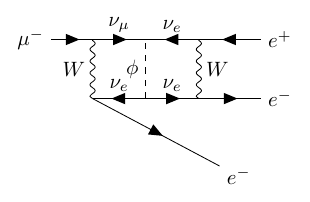}
\vspace{-10pt}
\caption{Feynman diagram for the 2-loop decay $\mu\to eee$ induced by $\phi$ via the couplings $h_{ee,\,e\mu}$ in Eq.~(\ref{eq:LNVLagrangian}).}
\label{fig:mu3e}
\end{figure}

If the active neutrinos are Majorana fermions, the couplings of $\phi$ could induce the rare LFV decay $\mu \to eee$ at the 2-loop order via the couplings $h_{ee,\, e\mu}$, and the Feynman diagram is shown in Fig.~\ref{fig:mu3e}~\cite{Berryman:2018ogk}. The corresponding BR is roughly
\begin{equation}
{\rm BR}(\mu\to eee) \sim \frac{G_F^2 m_W^4}{256 \pi^4} |h_{ee}h_{e\mu}|^2 \,.
\end{equation}
Although no neutrino mass suppression appears in the amplitude, the process is heavily suppressed by the 2-loop factor. 
The current limit on ${\rm BR}(\mu \to eee)$ from SINDRUM~\cite{SINDRUM:1987nra} is $1\times10^{-12}$, and the resulting constraint is $|h_{ee}h_{e\mu}| \lesssim {\cal O}(10^{-2})$~\cite{Berryman:2018ogk}, which is weaker than that from the $G_F$ measurements (cf. Fig.~\ref{fig:limit2}). The sensitivity in this channel can be improved by up to four orders of magnitude at Mu3e, and the resulting constraint on the coupling $|h_{\alpha\beta}|$ will be strengthened by one order of magnitude~\cite{Mu3e:2020gyw}.

\subsection{$\mu \to e$ conversion in nuclei}

\begin{figure}[t!]
\centering
\includegraphics[width=0.49\linewidth]{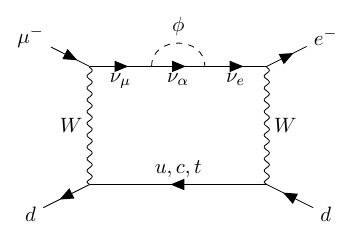}
\includegraphics[width=0.49\linewidth]{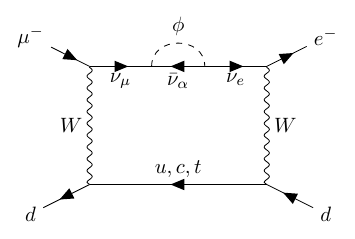}
\vspace{-10pt}
\caption{Representative Feynman diagrams for the 2-loop $\mu\to e$ conversion in nuclei for Dirac (left) and Majorana (right) neutrinos, induced by the couplings $g_{\alpha\beta}$ and $h_{\alpha\beta}$ of $\phi$ in Eq.~(\ref{eq:Lagrangian}), respectively. 
}
\label{fig:mu-e_conv}
\end{figure}

The neutrinophilic scalar $\phi$ can also induce coherent $\mu\to e$ conversion in nuclei at the 2-loop order, with the scalar--neutrino loop inserted into the neutrino propagator of the $W$-mediated box diagram. The representative Feynman diagrams with external $d$ quark legs are presented in Fig.~\ref{fig:mu-e_conv}, with the left and right diagrams again for the Dirac and Majorana neutrinos, respectively. The $\mu\to e$ conversion can also occur for the external $u$ quark legs, the $d,\,s,\,b$ quark mediators and the opposite quark flow. Using the standard normalization of the conversion rate in
Ref.~\cite{Alonso:2012ji}, an order of magnitude estimate gives
\begin{equation}
\mathrm{R}(\mu N \rightarrow e N)
\sim
\frac{18G_F^4m_W^4m_\mu^5}{(4\pi)^8\Gamma_{\rm capt}}
g_{\rm eff}^4
\left|V^{(p)}+V^{(n)}\right|^2 ,
\end{equation}
where $\Gamma_{\rm capt}$ is the total muon capture rate of the target nucleus, and $V^{(p)}$ and $V^{(n)}$ denote the corresponding proton and neutron nuclear overlap integrals, respectively. 
The current most stringent constraint  of $\text{R}(\mu\, {\rm Au}\to e\,{\rm Au})<7\times10^{-13}$ is from the SINDRUM II experiment using the gold target~\cite{SINDRUMII:2006dvw}. With the representative values of $V^{(p)}\simeq0.0974$, $V^{(n)}\simeq0.146$, and $\Gamma_{\rm capt}\simeq13.07\times10^{6}\ {\rm s}^{-1}$, the resulting limit on the coupling is $g_{\rm eff} \lesssim 0.4$. It is expected that the reaches of $g_{\rm eff}$ could be significantly improved at future $\mu\to e$ conversion experiments such as Mu2e~\cite{Mu2e:2014fns} and COMET~\cite{COMET:2018auw}.

\section{Conclusion}
\label{sec:conclusion}

In this paper, we have performed a comprehensive study of the effects of the neutrinophilic scalar $\phi$ on muon decays, assuming the SM neutrinos are either Dirac or Majorana particles. We have considered all the possible exotic muon decay channels that are induced by $\phi$ at the tree-level, 1-loop and 2-loop orders. In particular, we focus on the following processes: (i) the exotic three-body decay
$\mu\to e\phi\phi$ and the $\phi$-induced 2-loop decay $\mu \to e \gamma$ induced by the coupling $g_{\rm eff}$ in Eq.~(\ref{eq:geff}) for Dirac neutrinos (and also for Majorana neutrinos); (ii) the $\phi$-induced 1-loop corrections to the SM muon decay
$\mu\to e\nu\bar{\nu}$, and the associated real emission channels
$\mu^- \to e^-\nu_\mu\nu_\mu\phi,\, e^- \bar\nu_e \bar\nu_e \phi$ for Majorana neutrinos, induced by the LFV coupling $h_{e\mu}$. There are IR divergences in the limit of $m_\phi \to 0$, which cancel out when the corresponding partial widths are summed up and yield an IR safe inclusive muon decay width. There are also some other muon relevant processes induced by $\phi$, which are, however, highly suppressed by the tiny neutrino masses and/or the 2-loop factor. 

We derive constraints on the couplings of $\phi$ to neutrinos from the precise measurements of the Fermi constant $G_F$ and the experimental bounds on ${\rm BR}(\mu \to e \gamma)$, and compare them with the existing limits from charged 
meson decays, EWPD, the IceCube high-energy neutrino events, CMB and BBN. Here are the main results.
\begin{itemize}
    \item Although the $\phi$-induced LFV decay $\mu\to e\gamma$ arises at the 2-loop level, it provides the most stringent constraints on the coupling $g_{\rm eff} \lesssim 0.064$ if the scalar mass $m_\phi \gtrsim 8.7$ MeV, while the limits from $G_F$ measurements are much weaker (cf. Fig.~\ref{fig:limit1}).
    
    \item For the 1-loop corrections of $\phi$ to the SM muon decay and the four-body decays $\mu^- \to e^- \nu_\mu \nu_\mu \phi,\, e^- \bar\nu_e \bar\nu_e \phi$, the partial widths are proportional to $|h_{e\mu}|^2$, and the $G_F$ measurements have excluded large regions of parameter space, with the coupling $|h_{e\mu}|$ down to 0.027. As a result of the radiative effects of $\phi$ on the SM muon decay, the scalar mass can be probed far above the muon mass, up to roughly 5.9 TeV. In particular, the $G_F$ measurements provide the most stringent constraint on $m_\phi$ and $|h_{e\mu}|$ for scalar mass above roughly ${\cal O}({\rm GeV})$, as seen in Fig.~\ref{fig:limit2}.
\end{itemize}

Although $G_F$ is one of the most precisely measured parameters of the SM, there seems to be some tension among the determinations from muon decay, CKM unitarity and EW fit, which could be alleviated by the exotic muon decays such as the 1-loop corrections of $\phi$ to the SM muon decay and the four-body decays $\mu^- \to e^- \nu_\mu \nu_\mu \phi,\, e^- \bar\nu_e \bar\nu_e \phi$ (cf. Fig.~\ref{fig:limit2}). However, such tensions need to be confirmed (or falsified) in future high-precision data. More importantly, we emphasize that a global analysis of the muon decay data, including both the muon lifetime and the Michel parameters, and other high-precision EW data could further improve the constraints on the neutrinophilic scalars.

\acknowledgments

The authors would like to thank Patrick D. Bolton, Shufang Qiang and Peiwen Wu for useful discussions. We also thank Zeren Simon Wang for the valuable comments on the manuscript. Z.Z. is grateful to IPPP, Durham for hospitality while this work was being finalized. This work is supported by the National Natural Science Foundation of China (NSFC) under Grant No. 12175039. This work is also supported by State Key Laboratory of Dark Matter Physics and the Fundamental Research Funds for the Central Universities.

\bibliography{biblio}
\bibliographystyle{JHEP}

\end{document}